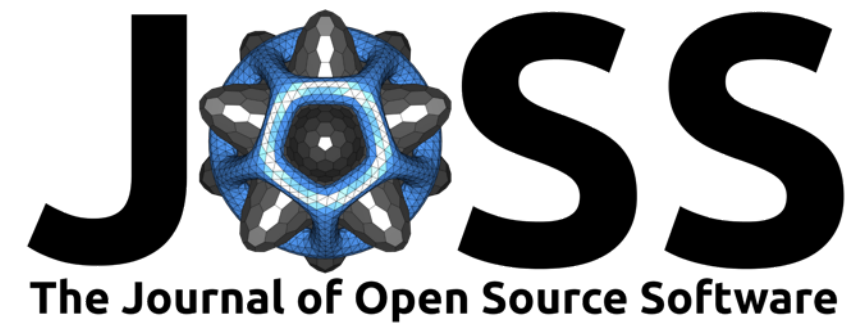

# ForMoSA: Forward Modeling tool for Spectral Analysis

**ForMoSA Collaboration**[1,2,3,4,5,6,7,8,9,10*], **Simon Petrus**[1,2,3,4*], **Paulina Palma-Bifani**[5,6*], **Matthieu Ravet**[6,4,7*], **Allan Denis**[8*], **Bhavesh Rajpoot**[7,9*], **Mickaël Bonnefoy**[4], **Gaël Chauvin**[7,6], **Arthur Vigan**[8], **Alice Radcliffe**[5], **Pablo Requeijo**[5], and **Kevin Hoy**[2,3,10]

**1** NASA-Goddard Space Flight Center, Greenbelt, MD 20771, USA **2** Instituto de Estudios Astrofísicos, Facultad de Ingeniería y Ciencias, Uni. Diego Portales, Av. Ejército 441, Santiago, Chile **3** Millennium Nucleus on Young Exoplanets and their Moons **4** Univ. Grenoble Alpes, CNRS, IPAG, F-38000 Grenoble, France **5** LIRA, Observatoire de Paris, Université PSL, Sorbonne Université, Université de Paris, 5 place Jules Janssen, 92195 Meudon, France **6** Laboratoire J. L. Lagrange, Université Côte d'Azur, Observatoire de la Côte d'Azur, CNRS, 06304 Nice, France **7** Max-Planck-Institut für Astronomie, Königstuhl 17, 69117 Heidelberg, Germany **8** Aix Marseille Univ, CNRS, CNES, LAM, Marseille, France **9** Department of Physics and Astronomy, Heidelberg University, Im Neuenheimer Feld 226, D-69120 Heidelberg, Germany **10** European Southern Observatory, Alonso de Cordova 3107, Vitacura, Santiago, Chile ***** These authors contributed equally.

## Summary

**`ForMoSA` (FORward MOdeling tool for Spectral Analysis)** is an open-source `Python` package to fit spectroscopic and photometric observations using a Bayesian framework. It can utilize different self-consistent atmospheric models to perform robust parameter space exploration. It has been mainly designed for fitting directly imaged young planetary-mass brown dwarfs and exoplanets. The developments within **`ForMoSA`** are supported by an international collaboration of several laboratories in France (IPAG, LIRA, LAM, and Lagrange), Germany (MPIA), USA (NASA Goddard), and Chile (FCLA, Universidad Diego Portales, and Universidad de Chile). The evolution of the code and the growing interest from the scientific community has led to the need for this dedicated publication alongside the release of **`ForMoSA` v2.0**, which has been refactored into a class architecture with user-friendly features and extended functionalities.

## Statement of need

Recent advances in ground- and space-based observatories now enable routine, high-quality observations of exoplanet atmospheres across a wide range of wavelengths and resolutions. In practice, **`ForMoSA`** was designed to bridge the gap between these observations and atmospheric models by providing a Bayesian framework to robustly compare the two. **`ForMoSA`** adopts a forward-modeling approach based on exploring through a nested sampling (Skilling, 2004) the parameter space of pre-computed, self-consistent atmospheric model grids.

The code **`ForMoSA`** addresses the following needs:

- **Data modularity:** It allows the simultaneous analysis of diverse datasets, even when they overlap in wavelength coverage. This includes multiple datasets taken at different times and with different setups.
- **Computational efficiency:** It leverages `xarray`[1], enabling fast interpolation and manipulation of large model grids.

[1] `xarray` documentation: https://docs.xarray.dev/en/stable

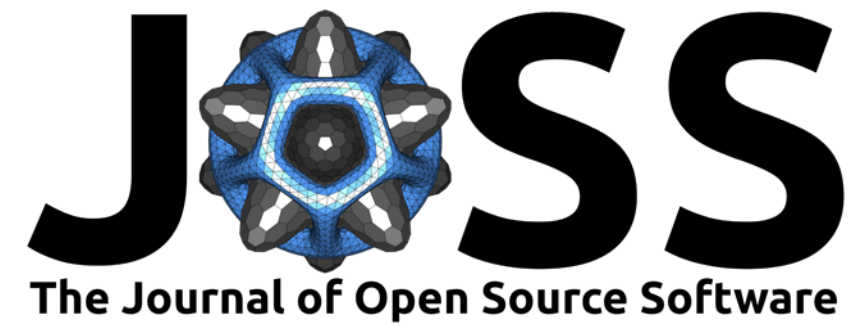

- **Flexible likelihood mappings:** Users can choose from various likelihood metrics to account for specific observational uncertainties (Ruffio et al., 2019).
- **Extensive parameterization:** Beyond the atmospheric model parameters, a comprehensive suite of extra-grid parameters, such as radial velocity, rotational broadening, and scaling factors, can be incorporated into the fit.
- **High-Contrast module:** It incorporates the method of (Landman et al., 2024), allowing users to include reference stellar, atmospheric transmission, and systematics spectra to accurately fit contrast-limited data.

# State of the field

The landscape of exoplanetary atmospheric modeling is broadly divided into two categories: "free retrieval" frameworks, which compute radiative transfer on the fly to fit parameterized atmospheric structures, and "forward modeling" codes, which interpolate pre-computed grids of self-consistent physical models.

The codes in the first category are **CHIMERA** (Line et al., 2013), **petitRADTRANS** (Mollière et al., 2019), **NEMESIS** (Irwin et al., 2008), **CROCODILE** (Hayoz et al., 2023), and **Pyrat Bay** (Cubillos & Blecic, 2021). These tools are very flexible but also computationaly expensive. Other specialized tools can handle specific niche problems, such as opacity management and radiative transfer calculations (**Exo_k**, Leconte, 2021), high-resolution autodifferentiation (**ExoJAX**, Kawahara et al., 2022), 1D disequilibrium chemistry (**HOMER**, Himes et al., 2022), or models for rocky planet emission spectra (**rfast**, Robinson, 2023).

**ForMoSA** complement these on-the-fly radiative transfer or free-retrieval codes by offering a faster, grid-interpolated alternative for parameter estimation. Our code's closest existing counterparts are **species** (Stolker et al., 2020) and **SEDA** (Suarez et al., 2025), which also facilitate Bayesian inference between spectro-photometric observations and pre-computed atmospheric grids. **species** is a generalized, all-in-one toolkit for direct imaging that encompasses everything from flux calibration to color-magnitude diagrams. **SEDA** is more similar to **ForMoSA**, but it was only introduced early 2025 to the community, and has not yet been widely adopted. We chose to build **ForMoSA** as a standalone package to provide a highly specialized, modular inversion framework expressly optimized for data heterogeneity and statistical flexibility. Integrating these new capabilities into existing tools would have required fundamental rewriting, leading to a new code.

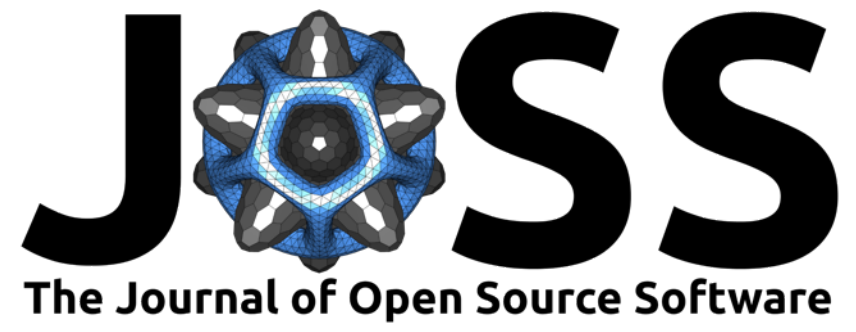

## Software design

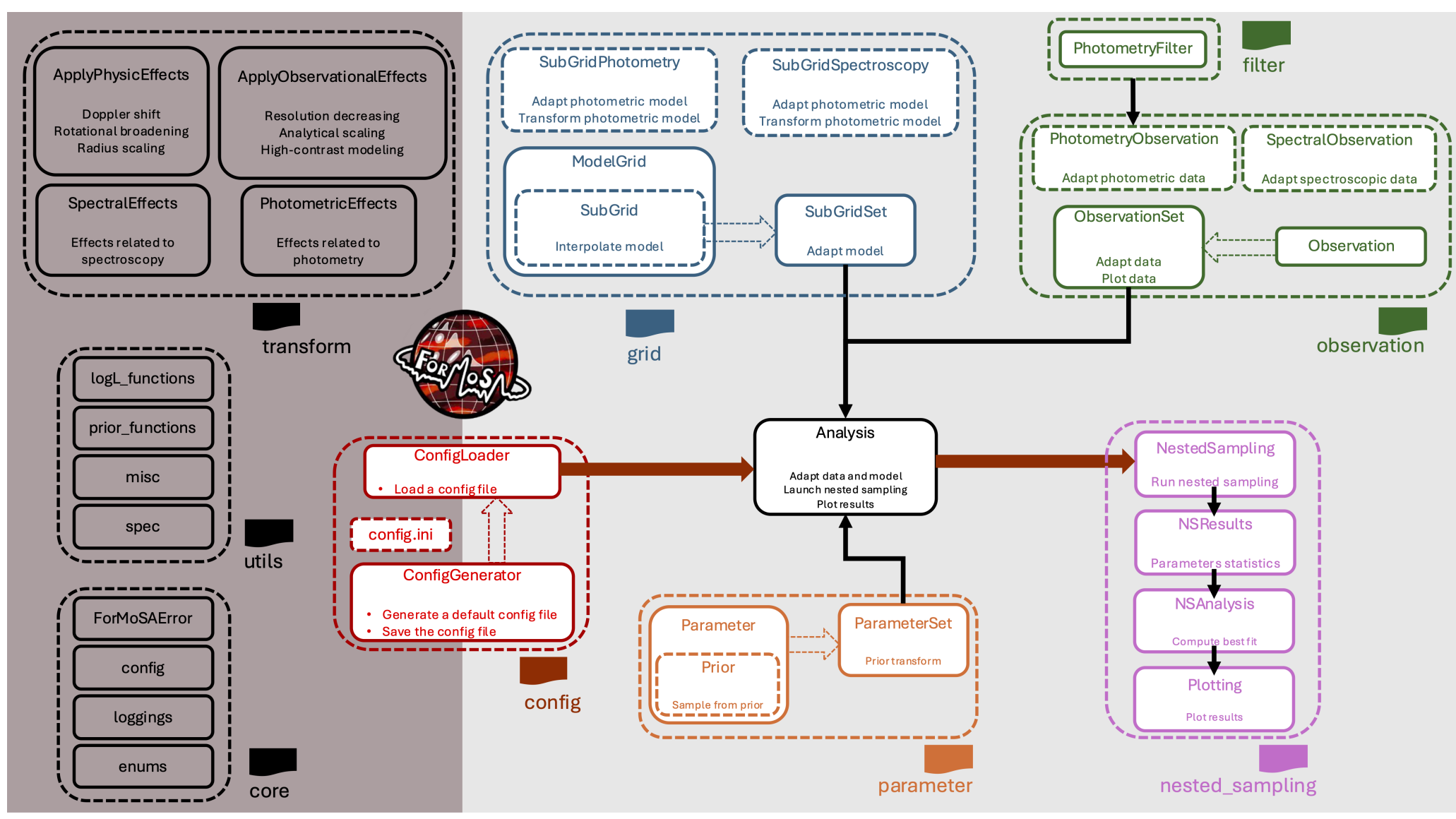

**Figure 1: ForMoSA** workflow diagram. The shaded gray area on the right represents the core functionalities required to run the Nested Sampling. The left dark-gray area represents utility and support functions. The small boxes contain the modules, and the dashed boxes represent the contents of each folder.

The Python-based code of **ForMoSA** is organized into several modules, as shown in Figure 1. In practice, **ForMoSA** v2.0 has a class-based architecture, with some main modules (config, observation, filter, grid, parameter, nested_sampling), and support modules (core, utils, and transform). Because **ForMoSA** is an end-to-end Bayesian tool that handles all stages of the analysis, its installation requires several `Python` packages, as well as downloading the model grids needed for the fit. All information required for installation, access to available grids, and the tutorials material is available at the **ForMoSA** documentation[2].

In summary, to start an analysis, the user must first provide an atmospheric model grid, the observations, and a `.ini` configuration file. This `.ini` file contains all of **ForMoSA**'s input parameters, including the paths, grid adaptation keys, nested sampling keys, and the free parameters priors configuration. **ForMoSA**'s main modules are centralized through a main Analysis class, which creates and handles sub-grids of atmospheric models tailored to each observation, before launching the nested sampling algorithm. At each iteration of the nested sampling, random parameter values are drawn from the prior distributions. The sampled parameters are used to interpolate model spectra from the sub-grids, before comparing them to each observation by computing the log-likelihood function. For a given run, the data, sub-grids, parameters, and results are automatically saved to paths specified by the user, and can be easily recovered for subsequent analysis and visualized through the Plotting class.

## Research impact statement

Initiated in 2020, **ForMoSA** has become a highly adaptable tool for the atmospheric characterization of substellar objects. Its computational efficiency and compatibility with diverse physical models have enabled the standard analysis of targets across a wide range of spectral resolutions, from broadband photometry to high-resolution spectroscopy.

[2] **ForMoSA** documentation: https://ForMoSA.readthedocs.io/en/latest/index.html

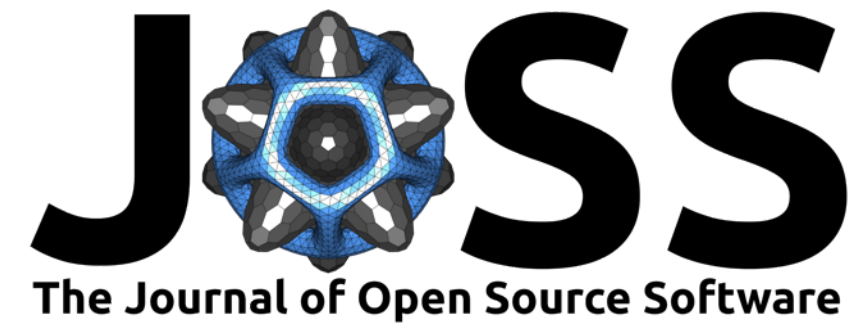

A major strength of **ForMoSA** lies in its capacity to jointly fit multi-instrument, heterogeneous datasets to maximize wavelength coverage. Because combining data from diverse instruments often introduces biases, **ForMoSA** features the dedicated `MOSAIC` module. This framework utilizes customized likelihood scaling and parametric covariance models to mitigate inter-calibration offsets and complex noise structures, which can be tuned for each specific target and dataset. Altogether, these features allow for the accurate characterization and publication of atmospheric properties of various benchmark companions such as HIP 65426 (Carter et al., 2023; Petrus et al., 2021), AB Pic b (Palma-Bifani et al., 2023), VHS 1256 b (Petrus et al., 2023, 2024; Radcliffe et al., 2026), AF Lep b (Palma-Bifani et al., 2024), $\beta$ Pic b (Houllé et al., 2025; Ravet et al., 2025), YSES 1 b and c (Hoch et al., 2025), and COCONUTS-2 b (Ravet et al., 2026). Furthermore, **ForMoSA**'s speed allows for the homogeneous analysis of large spectral libraries, such as those of brown dwarfs and planetary-mass companions (Palma-Bifani et al., 2025; Petrus et al., 2020, 2025), resulting in the identification of parameter trends and structural systematics in atmospheric models. Finally, **ForMoSA** also includes a high-contrast, high-resolution (HCHR) module applied to contrast-limited, stellar-contaminated data (e.g. VLT/HiRISE) to extract precise radial velocities and detailed orbital constraints (Denis et al., 2025, 2026).

**ForMoSA** is an active project, and its scientific production is growing rapidly, therefore a list of peer-reviewed publications is maintained on our NASA ADS public library[3]. Our code will continue to evolve over time, with new functionalities being added to meet the requirements of data from future observatories (ELTs, HWO, etc.), updates to the state-of-the-art atmospheric models, and the parametrization of additional physics to better fit the data (e.g., disks, extinction laws, multiple columns, time variability). For this reason, the official documentation is regularly updated at `ReadTheDocs`, and should be considered the reference description of the code.

## AI usage disclosure

Our team fully designed and authored the core architecture, logic, and implementation of the code without AI generation. We used AI strictly as an auxiliary tool for debugging and managing documentation. Additionally, we utilized Gemini to refine the English and Gemini coupled with Antigravity to format and compile the main text and the bibliography.

## Acknowledgements

The authors express their sincere thanks to the Code/Astro Workshop[4], which provided the foundational training necessary to transform **ForMoSA** into a professional, open-source `Python` package. We gratefully acknowledge the funding and support for the ForM-X workshops held in Nice (2023), Heidelberg (2024/2025), and Grenoble (2025). These collaborative sessions were instrumental in the development and refinement of the code. We also thank the various laboratories and institutions, especially IPAG, Lagrange, and MPIA, for their continued support. Furthermore, this work has been supported by the French National Research Agency (ANR) through the MIRAGES project (PI: A. Vigan, ANR-20-CE31-0017).

## Appendix A: Performance

The computational cost of **ForMoSA** is primarily driven by the number of forward model evaluations required for the nested sampling algorithm to converge. While absolute execution time (inversion time) depends on the specific hardware, Figure 2 provides a comparative analysis to guide users in defining the configuration file. The total inversion time is a multi-dimensional function depending on several factors, including the spectral resolution ($R_\lambda$), the signal-to-noise ratio (S/N), the number of live points, the dimensionality of the parameter space, and the machine used.

[3] NASA ADS Library: https://ui.adsabs.harvard.edu/user/libraries/PekELjOGR4yl3XOnGwOAng
[4] Code/Astro Workshop: https://semaphorep.github.io/codeastro/

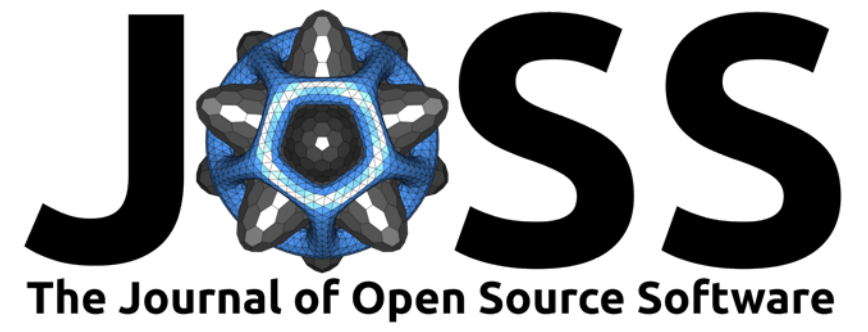

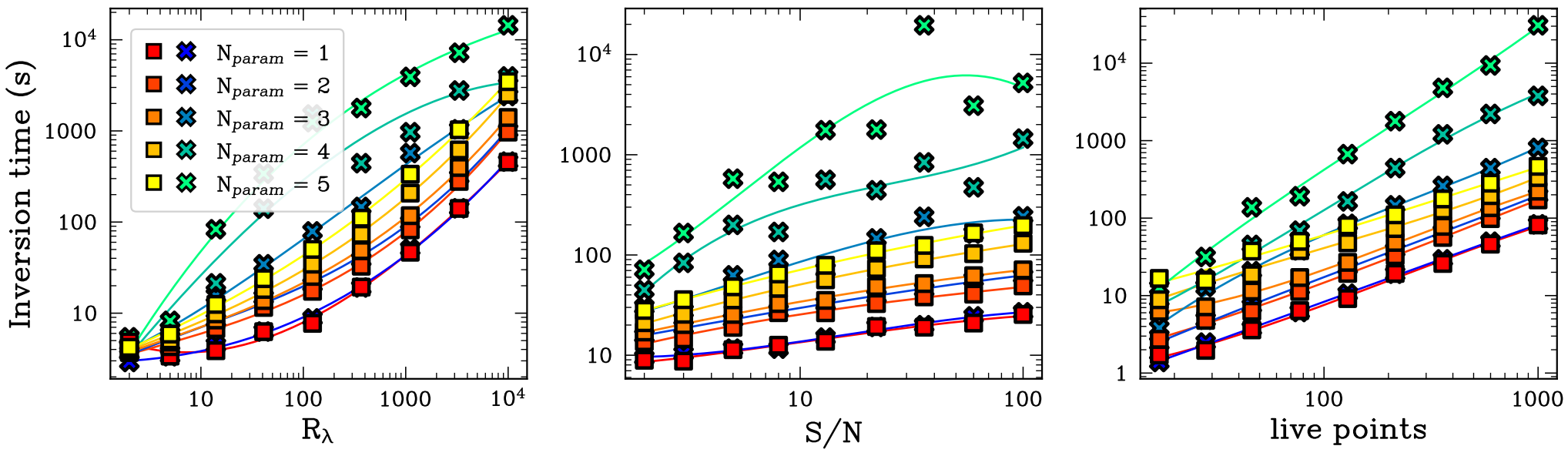

**Figure 2:** Performance comparison between the nested sampling algorithms `PyMultiNest` (squares) and `Nestle` (crosses). Different colors show the number of free parameters used (from 1 to 5). From left to right: Inversion time as a function of spectral resolution ($R_\lambda$), signal-to-noise (S/N), and number of live points. The default setup is $R_\lambda$ = 368, S/N = 22, and 215 live points.

In the left panel of Figure 2, we observe that the main driver of computational cost is indeed the spectral resolution. As the spectral resolution increases, the number of data points also increases, and the number of points to be compared in the likelihood function grows, directly increasing the computational cost of each forward model call. In the middle panel, we observe the dependence on the S/N. Here, noise is applied using the following formula:

$$\vec{n} \sim \mathcal{N}\left(0, \sigma^2\right), \quad \text{with} \quad \sigma = \frac{\mathrm{mean}(d)}{\mathrm{S/N}},$$

where $d$ is the synthetic flux. As the S/N increases, the error bars become smaller, which leads to a more sharply peaked likelihood function. This tighter posterior distribution requires the sampler to explore the parameter space with finer precision and typically increases the number of likelihood evaluations, thereby lengthening the total inversion time. In the right panel, we observe that the inversion time scales approximately linearly with the number of live points. The different colors in Figure 2 indicate how the dimensionality of the parameter space (number of free parameters) strongly drives the inversion time. As the complexity of the model increases, the “volume” of the prior space grows, requiring more iterations to isolate the high-likelihood regions. In addition, we also provide a comparison between `PyMultiNest` (squares) and `nestle` (crosses). While `nestle` is a convenient pure-Python implementation, `PyMultiNest` (leveraging the Fortran `MultiNest` library) generally demonstrates superior scaling and efficiency.

To optimize the integration time of a **`ForMoSA`** inversion, we advise users to:

- Start low and run initial tests with a reduced number of live points (<100) to verify the model setup.
- Use a spectral resolution that matches the physical information content of the data; over-sampling the spectrum increases the inversion time without improving accuracy.
- Select the right algorithm for the inversion. For high-dimensional fits (>5 parameters), `PyMultiNest` is the recommended default.

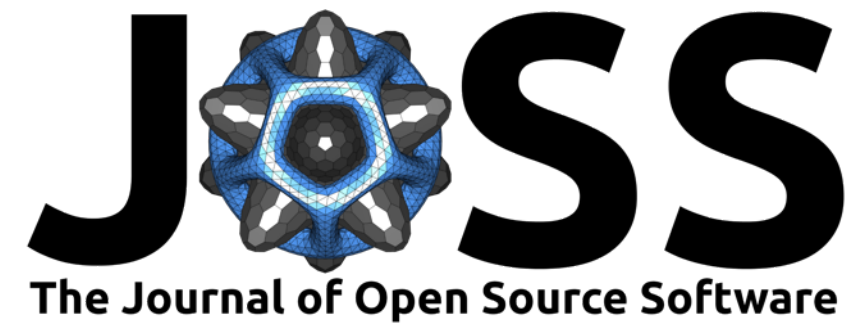

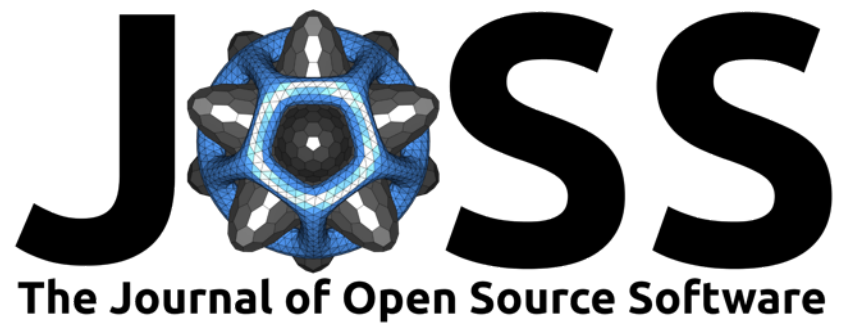

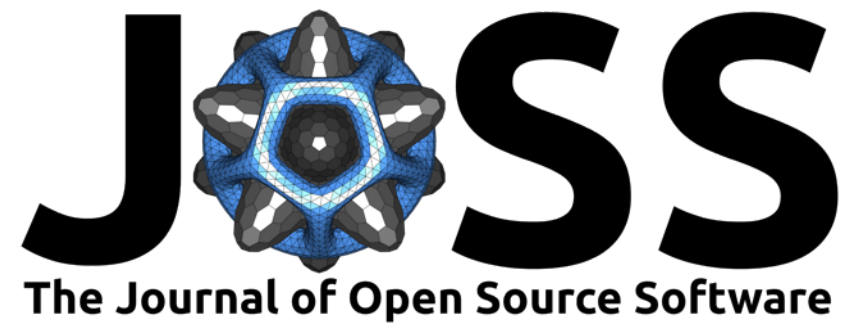